\documentclass[aps,prl,floatfix,twocolumn,superscriptaddress]{revtex4-1}
\usepackage{amsmath,amssymb}
\usepackage{graphicx}
\usepackage{epsfig}
\usepackage{psfrag}
\usepackage[usenames]{color}
\usepackage{xcolor}
\usepackage{hyperref}
\usepackage{soul,color}
\usepackage{tabu}
\usepackage{multirow}
\hypersetup{colorlinks=true,citecolor={blue},linkcolor={blue},urlcolor={blue}}

\begin{document}


\title{Bose--Einstein condensate-mediated superconductivity in graphene}














\author{Meng~Sun}
\affiliation{Center for Theoretical Physics of Complex Systems, Institute for Basic Science (IBS), Daejeon 34126, Korea}
\affiliation{Basic Science Program, Korea University of Science and Technology (UST), Daejeon 34113, Korea}

\author{A.~V.~Parafilo}
\affiliation{Center for Theoretical Physics of Complex Systems, Institute for Basic Science (IBS), Daejeon 34126, Korea}

\author{K.~H.~A.~Villegas}
\affiliation{Center for Theoretical Physics of Complex Systems, Institute for Basic Science (IBS), Daejeon 34126, Korea}
\affiliation{Division of Physics and Applied Physics, Nanyang Technological University, 637371 Singapore, Singapore}

\author{V.~M.~Kovalev}
\affiliation{A.~V.~Rzhanov Institute of Semiconductor Physics, Siberian Branch of Russian Academy of Sciences, Novosibirsk 630090, Russia}
\affiliation{Novosibirsk State Technical University, Novosibirsk 630073, Russia}

\author{I.~G.~Savenko}
\affiliation{Center for Theoretical Physics of Complex Systems, Institute for Basic Science (IBS), Daejeon 34126, Korea}
\affiliation{A.~V.~Rzhanov Institute of Semiconductor Physics, Siberian Branch of Russian Academy of Sciences, Novosibirsk 630090, Russia}

\date{\today}


\begin{abstract}
We propose a mechanism for robust BCS-like 
superconductivity in graphene placed in the vicinity of a Bose--Einstein condensate. Electrons in the graphene interact with the excitations above the condensate, called Bogoliubov quasiparticles (or bogolons). 
It turns out that bogolon-pair-mediated interaction allows us to surpass the long-standing problem of the vanishing density of states of particles with a linear spectrum. This results in a dramatic enhancement of the superconducting properties of graphene while keeping its relativistic dispersion. 
We study the behavior of the superconducting gap and calculate critical temperatures in cases with single-bogolon and bogolon-pair-mediated pairing processes, accounting for the complex band structure of graphene.
\textcolor{black}{We also compare the critical temperature of the superconducting transition with the BKT temperature.}
\end{abstract}	

\maketitle


Graphene is conventionally accepted as a two-dimensional (2D) material~\cite{Novoselov10451} with extremely high conductivity~\cite{Novoselov666, Castro-Neto:2009aa}.
Its chemical potential can be controlled by an external electric field; 
electrons and holes in graphene represent massless relativistic particles~\cite{MorozovNature} described by the 2D Dirac equation, 
which opens perspectives for outstanding transport characteristics and allows for the study of the interplay between relativity and superconductivity~\cite{Beenakker2008}. 
One interesting consequence of this interplay is the specular Andreev reflection in graphene--superconductor junctions~\cite{Beenakker2006}, which is not typical for normal metal--superconductor junctions where instead retroreflection takes place. 
When graphene is deposited on a substrate, 
it can adopt the properties of the latter such as ferromagnetism or superconductivity via the proximity effect~\cite{Heersche2007, PhysRevB.77.184507}, 
even though neither superconductivity nor ferromagnetism belong to the set of intrinsic properties of graphene.
All this makes hybrid graphene-based structures, such as graphene--superconductor interfaces, an intense topic of research that can broadly be called \textit{mesoscopic transport in graphene}. 
In particular, it brings in the term \textit{mesoscopic superconductivity}~\cite{Atienza2014}.


Why is bare graphene not intrinsically superconducting?
The primary reasons are the absence of electron--electron screening at small electron densities and the smallness of the electron density of states~\cite{CastroNetoPhysRevLett.98.146801}, which is linear in energy and thus vanishes at the Dirac point.
 %
As a consequence, electron--phonon interaction in graphene is strongly suppressed~\cite{Marchenko2018}. 
And since the Bardeen--Cooper--Schrieffer (BCS) electron pairing below the transition temperature $T_c$ involves basically the same matrix elements of electron--phonon interaction as the scattering matrix elements above $T_c$~\cite{Laussy:2010aa, Cotleifmmode-telse-tfi:2016aa, OurPRLBG, Skopelitis:2018aa}, there might be no BCS superconductivity in graphene other than that which is induced.
The bigger the matrix elements of the electron--phonon interaction are, the larger the gap opens and the larger $T_c$~\cite{Mahan} becomes, which are the benchmarks of robust superconductivity.
In other words, ``good'' conductors such as gold, copper, and graphene are ``bad'' superconductors.

Recently, techniques have been reported to turn multi-layer graphene into a superconductor by twisting bilayer graphene~\cite{Caoa2018, Yankowitz2019} or depositing it on a SiC substrate~\cite{Marchenko2018}. 
Trilayer graphene under a vertical displacement field also exhibits superconductivity~\cite{Chen2019}.
All these approaches are targeted at increasing the electron density of states at the Fermi energy by building a flat band~\cite{Kopnin2011, Munoz2013, Bistritzer2011}, including the recent progress in photonic graphene~\cite{WangNature2020}. 
Unfortunately, doing so usually destroys the \textit{relativistic} aspect of the problem since the dispersion of graphene no longer remains linear~\cite{PhysRevB.86.155449}. 
\textcolor{black}{
In the meantime, pairing of particles with truly linear spectrum might result in the emergence of new collective modes or, for instance, different values of the critical magnetic fields; moreover, nontrivial (exotic) superconducting pairing states ~\cite{CastroNetoPhysRevLett.98.146801,ZHANG17401,exotic1,exotic2,exotic3} as well as the multiband superconductivity~\cite{RefEfetov} in hybrid systems can be studied.
}

In this Letter, we propose a non-conventional 
mechanism for 
electron--electron pairing interaction (below $T_c$) in graphene, beyond the acoustic phonons and impurity channels~\cite{Das-Sarma:2011aa}. 
We consider a hybrid system consisting of graphene and a 2D Bose--Einstein condensate (BEC)~\cite{Butov2017, Fogler2014, Berman2016, Kasprzak:2006aa}.
Excitations above the BEC, called Bogoliubov quasiparticles (or bogolons), possess properties of sound and play significant role in electron scattering processes.
This leads us to surmise that graphene might acquire strong superconducting (SC) properties below $T_c$ due to the bogolon-mediated, as opposed to the acoustic phonon-mediated, pairing of electrons.
We check this assumption and prove it valid.
In this way, one state of matter (Bose condensate) can induce another state of matter (SC condensate) in \textcolor{black}{neighboring} graphene, avoiding a twist and securing its relativistic dispersion.

\begin{figure}[!t]
\includegraphics[width=0.48\textwidth]{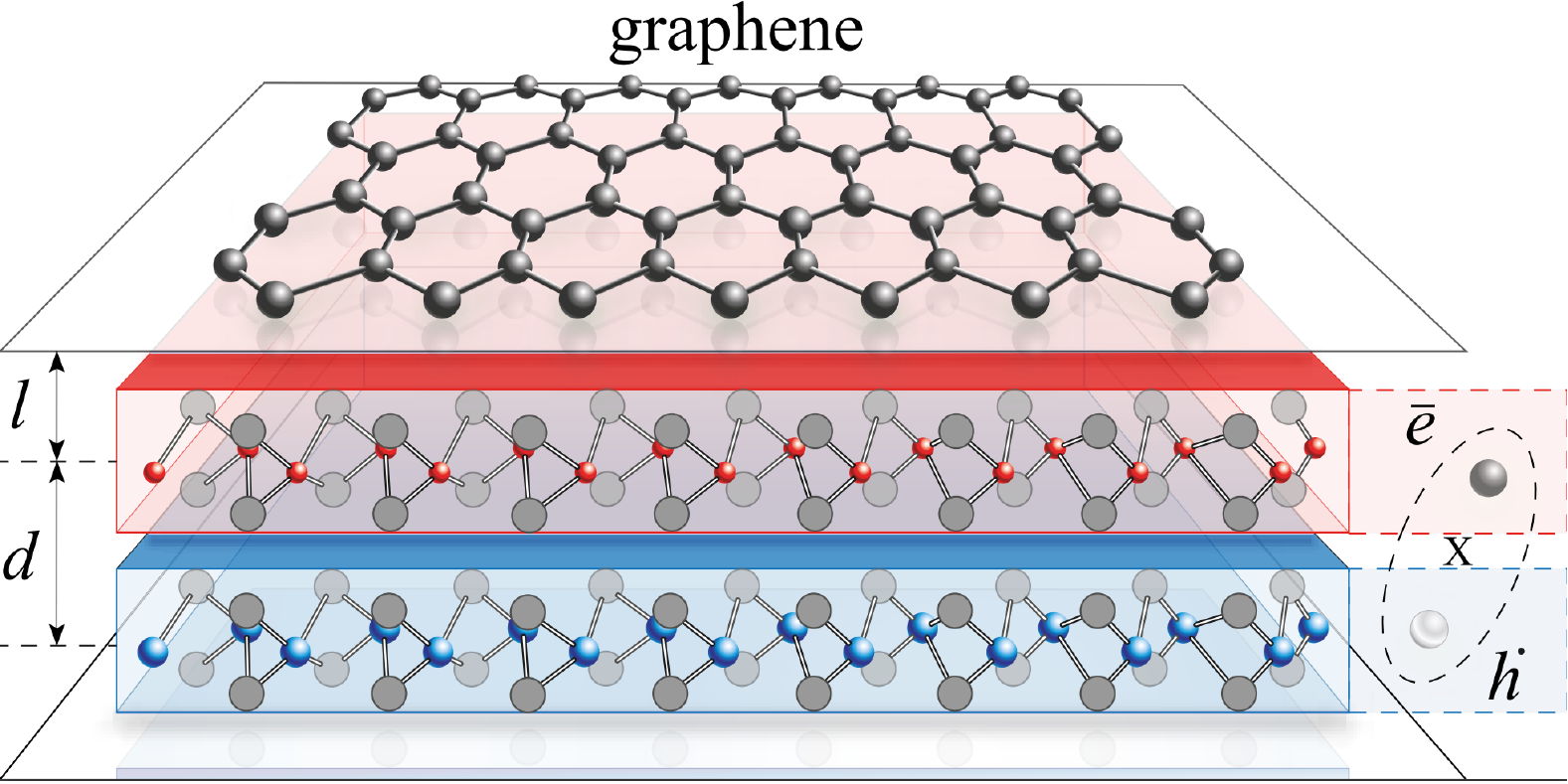}
\caption{System schematic: A hybrid system consisting of an electron gas in graphene (top layer) separated by distance $l$ from a two-dimensional Bose--Einstein condensate, represented by indirect excitons, where n-doped (red) and p-doped (blue) layers are separated by a spacer of thickness $d$.
The electrons in the graphene and the excitons are coupled by the Coulomb force.
}
\label{Fig1}
\end{figure}
%
%

Let us consider a system consisting of a 2D electron gas in graphene and two layers containing the indirect exciton gas made of MoS$_2$ (Fig.~\ref{Fig1}). 
The layers are spatially separated~\cite{[{We assume the absence of the tunneling between the electron and exciton layers despite the possible difference between the chemical potentials in two subsystems}]C1} and the particles are coupled by Coulomb interaction~\cite{Boev:2016aa, Matuszewski:2012aa}. 
%
%
%
%
%
This can be described by the Hamiltonian
\begin{equation}\label{eq.1}
{\cal H}=\int d\mathbf{r}\int d\mathbf{R}\Psi^\dag_\mathbf{r}\Psi_\mathbf{r}g\left(\mathbf{r}-\mathbf{R}\right)\Phi^\dag_\mathbf{R}\Phi_\mathbf{R},
\end{equation}
where $\Psi_\mathbf{r}$ and $\Phi_\mathbf{R}$ are the field operators of electrons and excitons, respectively, $g\left(\mathbf{r}-\mathbf{R}\right)$ is the Coulomb interaction strength, and $\mathbf{r}$ and $\mathbf{R}$ are the in-plane coordinates of the electron and exciton center-of-mass motion, respectively.
%
%
%

Furthermore, let us assume that the excitons are in the BEC phase, allowing us to use the model of a weakly interacting Bose gas and split $\Phi_\mathbf{R}=\sqrt{n_c}+\varphi_\mathbf{R}$, where $n_c$ is the condensate density and $\varphi_\mathbf{R}$ is the field operator of the excitations above the BEC.
Then Eq.~\eqref{eq.1} breaks into three terms,
\begin{eqnarray}
\nonumber
{\cal H}_{1}&=&\sqrt{n_c}\int d\mathbf{r}\Psi^\dag_\mathbf{r}\Psi_\mathbf{r} \int d\mathbf{R}g\left(\mathbf{r}-\mathbf{R}\right)\left[\varphi^\dag_\mathbf{R}+\varphi_\mathbf{R}\right],\\
\label{EqHam2}
{\cal H}_{2}&=&\int d\mathbf{r}\Psi^\dag_\mathbf{r}\Psi_\mathbf{r}\int d\mathbf{R}g(\mathbf{r}-\mathbf{R})\varphi^\dag_\mathbf{R}\varphi_\mathbf{R},
\end{eqnarray}
and ${\cal H}_3=gn_c\int d\mathbf{r}\Psi^\dag_\mathbf{r}\Psi_\mathbf{r}$, which gives a small correction to the Fermi energy $\mu$ and can usually be disregarded~\cite{Sun2018}.
\textcolor{black}{Here, we also disregard the effects related to finite lifetime of excitons thus leaving the consideration of their dynamics (including the pumping of the system) as well as the screening effects for the future study}~\cite{[{The screening effects may also play important role in multilayer systems~\cite{PhysRevB.89.060502}, in particular, bilayer graphene~\cite{PhysRevLett.110.146803}, preventing the formation of the BEC phase. 
However, it has recently been shown (in the framework of the RPA)~\cite{PhysRevB.96.174504} that the screening does not render any crucially detrimental effect on TMD bilayers, which we use in this Letter to lodge the excitonic condensate}]C6}.

We can now express the field operators of the bosonic excitations as a Fourier series of the linear combinations of bogolon annihilation (creation) operators $b_{\mathbf{q}}$($b^\dag_{\mathbf{q}}$),
\begin{eqnarray}
\label{EqExcField}
\varphi_\mathbf{R}
&=&\frac{1}{{L}}
\sum_{\mathbf{p}} e^{i\mathbf{p}\cdot\mathbf{R}}
(u_\mathbf{p}b_\mathbf{p}+v_{\mathbf{p}}b^\dagger_\mathbf{-p}),
\end{eqnarray}
where $L$ is a characteristic size and the Bogoliubov coefficients read~\cite{Giorgini:1998aa}
\begin{eqnarray}
\nonumber
&&u^2_{\mathbf{p}}=1+v^2_{\mathbf{p}}=\frac{1}{2}\left(1+\left[1+\left(\frac{Ms^2}{\omega_{\mathbf{p}}}\right)^2\right]^{1/2}\right),
\\
\nonumber
&&~~~~~u_{\mathbf{p}}v_{\mathbf{p}}=-\frac{Ms^2}{2\omega_{\mathbf{p}}},
\end{eqnarray}
with $M$ the exciton mass, $s=\sqrt{\kappa n_c/M}$ the sound velocity of the Bogoliubov quasiparticles, 
$\kappa=e_0^2d/2\epsilon_d$ 
the exciton--exciton interaction strength in the reciprocal space, $e_0$ the electron charge, $\epsilon_d=\epsilon\epsilon_0$, $\epsilon$ is the dielectric constant, $\epsilon_0$ is the vacuum permittivity, 
$\omega_\mathbf{p}=\hbar sp(1+p^2\xi_h^2)^{1/2}$ is the spectrum of bogolons with $p=|\mathbf{p}|$, and $\xi_h=\hbar/2Ms$ the healing length. %

Since the electron spectrum in graphene consists of two nonequivalent cones with minima at the Dirac points $\mathbf{K}$ and $\mathbf{K}'$, we can define the electron field operator as
\begin{eqnarray}
\label{EqElField}
\Psi_\mathbf{r}=\frac{1}{{L}}
\sum_{\mathbf{k},\sigma}\left(
e^{i(\mathbf{K}+\mathbf{k})\cdot\mathbf{r}}
c_{1,\mathbf{k},\sigma}+e^{i(\mathbf{K}'+\mathbf{k})\cdot\mathbf{r}}
c_{2,\mathbf{k},\sigma}\right),
\end{eqnarray}
where $c_{\alpha,\mathbf{k},\sigma}$ are electron annihilation operators in which $\alpha=1,2$ is the valley index and $\sigma=\uparrow,\downarrow$ is the electron spin projection. We disregard the spinor structure of the wave function in Eq.~(\ref{EqElField}) since we only consider the case of doped graphene with a Fermi energy sufficiently away from the Dirac points~\cite{ Castro-Neto:2009aa,RefEfetov}.
\textcolor{black}{We want to note, however, that the account of the pseudospin does not hinder the phenomena in question, as we explicitly show in the Supplementary Material~\cite{[{See the Supplementary Material at the [URL]}]SMBG}.}

%
From Eqs.~\eqref{EqHam2}--\eqref{EqElField}, we find
\begin{eqnarray}
\label{Eq1bexpr}
&&{\cal H}_1=
\sqrt{n_c}
\sum_{\bf k, p}\sum_{\alpha,\beta,\sigma}
\frac{g_{\bf p}^{\alpha\beta}}{L}
\Bigl[
(v_{\bf -p}+u_{\bf p})b_{\bf p}
\\
\nonumber
&&
~~~~~~~~~~~~~~~~~~~~+
(v_{\bf p}+u_{\bf -p})b^{\dag}_{-\bf p}
\Bigr]
c^{\dag}_{\alpha,\bf k+p,\sigma}c_{\beta,\bf k, \sigma},\\
\nonumber
&&{\cal H}_2=
\sum_{\bf k, p, q}\sum_{\alpha,\beta,\sigma}
\frac{g_{\bf p}^{\alpha\beta}}{L^2}
\left[
u_{\bf q-p}u_{\bf q}b^{\dag}_{\bf q-p}b_{\bf q}+u_{\bf q-p}v_{\bf q}b^{\dag}_{\bf q-p}b^{\dag}_{\bf -q}\right.\\
\nonumber
&&\left.
~~
+v_{\bf q-p}u_{\bf q}b_{\bf -q+p}b_{\bf q}
+v_{\bf q-p}v_{\bf q}b_{\bf -q+p}b^{\dag}_{\bf -q}
\right]
c^{\dag}_{\alpha,\bf k+p,\sigma}c_{\beta,\bf k,\sigma},
\end{eqnarray}
%
where $g_\mathbf{p}^{\alpha\beta}$ is
the
Fourier image of the electron--exciton interaction (we define its explicit form below). 
The terms ${\cal H}_1$ and ${\cal H}_2$ correspondingly describe single-bogolon (1b) and bogolon-pair (2b)-mediated processes.
Note that the 1b and 2b processes are of the same order with respect to the electron--exciton interaction strength $g_{\bf p}^{\alpha\beta}$.

\begin{figure}[!b]
    \centering
    \includegraphics[width=0.49\textwidth]{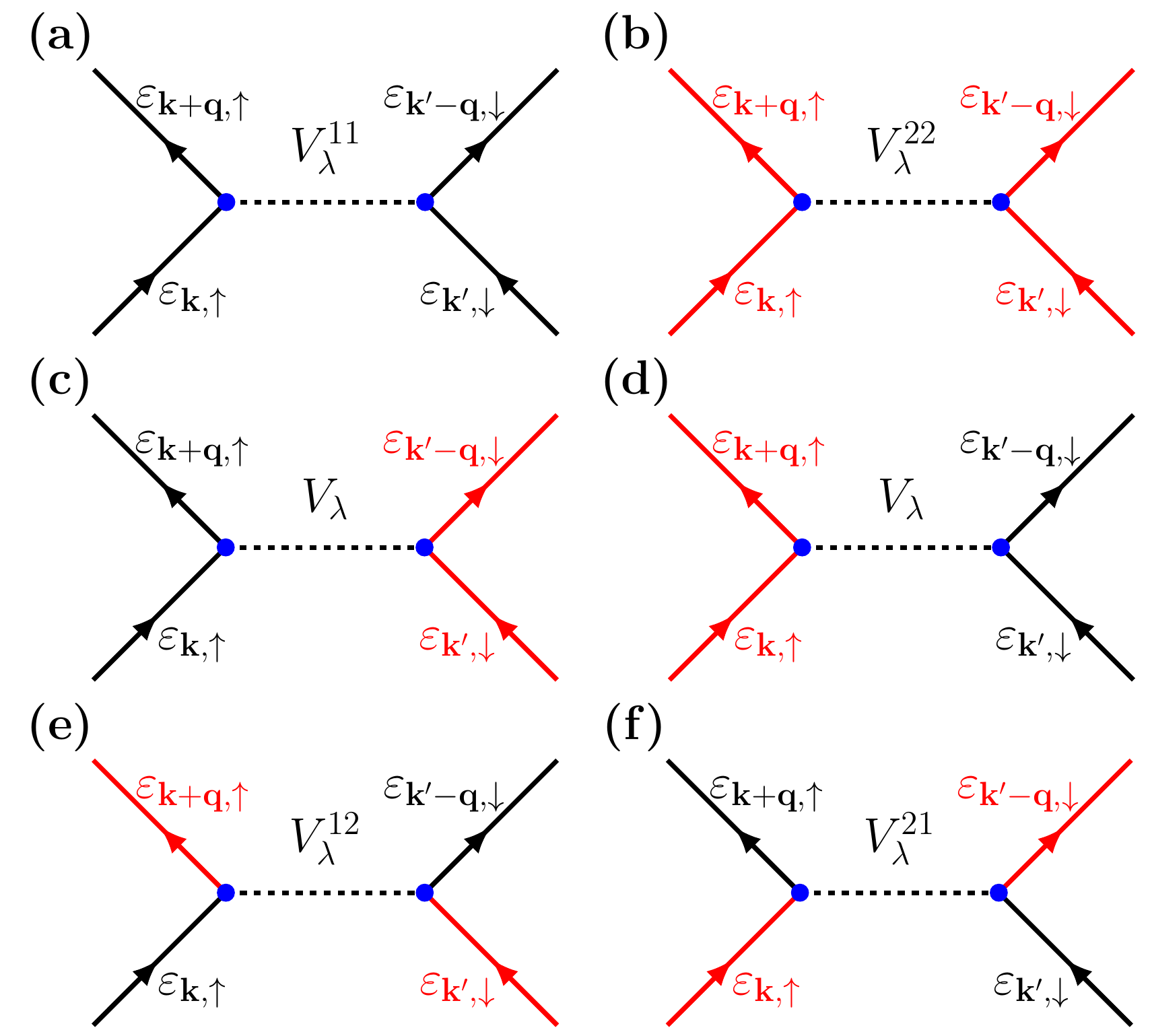}
    \caption{Feynman diagrams illustrating
    intravalley (a,b) and intervalley (c--f) pairings in accordance with the Hamiltonian~\eqref{effectiveinter}. Red and black lines describe electrons in the K and K' valleys, respectively.}
    \label{fig:feynman}
\end{figure}

First, we want to integrate out the bogolon degrees of freedom by using a standard procedure based on the Schriffer--Wollf transformation over the terms in the Hamiltonian~(\ref{Eq1bexpr}).
As the result, we find an effective Hamiltonian for 1b or 2b pairings ($\lambda=1\textrm{b}$, $2\textrm{b}$),
\begin{eqnarray}
\label{effectiveinter}
&&{\cal H}^{(\lambda)}_\textrm{eff}={\cal H}_0+
\\
\nonumber
&&~~+\sum_{\bf k,k',p}\sum_{\sigma,\sigma'}\sum_{\alpha,\beta}
\frac{V_{\lambda}^{\alpha\beta}(p)}{2L^2}
c^{\dag}_{\alpha,{\bf k+p},\sigma}c_{\beta,{\bf k},\sigma}c^{\dag}_{\alpha,{\bf k'-p},\sigma'}c_{\beta,{\bf k'},\sigma'}\\
\nonumber
&&~~+\sum_{\bf k,k',p}\sum_{\sigma,\sigma'}
\sum_{\alpha\neq\beta}
\frac{V_{\lambda}(p)}{2L^2}
c^{\dag}_{\alpha,{\bf k+p},\sigma}c_{\alpha,{\bf k},\sigma}c^{\dag}_{\beta,{\bf k'-p},\sigma'}c_{\beta,{\bf k'},\sigma'},
\end{eqnarray}
where ${\cal H}_0$ is the
kinetic energy of electrons in graphene, 
$V_{\lambda}^{11}(p),V_{\lambda}^{22}(p)=V_{\lambda} (p)$ are the matrix elements responsible for intravalley electron scattering, and $V_{\lambda}^{12(21)}(p)=V_{\lambda}(p\mp Q)$ are the intervalley scatterings with $Q=|\bf K - \bf K'|$ the momentum difference between the Dirac points.
The corresponding Feynman diagrams for two-electron scattering are plotted in Fig.~\ref{fig:feynman}.

After algebraic calculations~\cite{SMBG}, we find
\begin{eqnarray}
\label{EqEffPot1b}
V_{1b}(p)&=&-\frac{n_c}{Ms^2}g_p^2\equiv-\chi_{1},
\\
\label{EqEffPot2b}
V_{2b}(p)&=&
-\frac{\chi_{2}}{p}
\left(
1+
\frac{8}{\pi}
\int\limits_{L^{-1}}^{p/2 } 
\frac{N_qdq}{\sqrt{p^2-4q^2}}
\right),
\end{eqnarray}
%
where $\chi_{2}=M^2sg_p^2/4\hbar^3$, and $g_p=e^2_0 (1-e^{-pd})e^{-pl}/2\epsilon_d p$; $l$ is the distance between graphene and $n$-doped MoS$_2$ layer, while $d$ is the distances between $n$- and $p$-doped layers of the MoS$_2$, as shown in Fig.~\ref{Fig1}~\cite{[{
We see that the dependence of $g_p$ on $d$ and $l$ is exponential, and also that $g_p$ depends on the interlayer material. Existing experiments allow for a nanometer scale of the spacings~\cite{WangNature2019}.
The infrared cut-off $L^{-1}$ is necessary for the convergence. Its physical meaning is the absence of fluctuations with wavelengths larger than $L$. It can also be related to the critical temperature of BEC formation in a finite trap of length $L$~\cite{Bagnato1991}. A trapping is required~\cite{Butov2017} since there is no long-range order in infinite homogeneous 2D systems at finite temperatures~\cite{Hohenberg1967}}]C2}. 
The term $V_{1b}(p)$ corresponds to electron--electron pairing with an exchange of a single Bogoliubov excitation of the BEC, whereas $V_{2b}(p)$ describes electron pairing mediated by an exchange of a pair of bogolons. 
To derive these formulas, we followed the BCS approach and considered a constant attractive interaction between electrons with energies smaller than a cut-off, $\omega_b=\hbar s/\xi_h$, which appears by analogy with the Debye energy in BCS theory. 
Since $\xi_h\sim1/s$ and thus $\omega_b\sim s^2\sim n_c$, the typical energy scale of the attractive interaction can be controlled by the density of the condensate, $n_c$. 
In order to satisfy the applicability of the BCS theory, we also assume that {\color{black} $\omega_b\ll \hbar v_0/a \sim 4.6$ eV}, where $a$ is the interatomic distance in graphene~\cite{kopnin}. 

From a comparison of Eqs.~\eqref{EqEffPot1b} and~\eqref{EqEffPot2b}, we see that, in contrast to the 1b-mediated superconductivity, the strength of the 2b-mediated pairing potential contains an additional temperature-dependent term proportional to the Bose distribution of bogolons, $N_q\equiv [\exp(\hbar \omega_q/k_BT)-1]^{-1}$. 

Now, we are armed to proceed with the SC order parameter, whose structure is nontrivial due to the presence of two valleys~\cite{RefEfetov}. 
Indeed, we can distinguish between two SC gaps, namely diagonal and nondiagonal in the valley indices: $\Delta_{\lambda}^{\alpha\alpha}$ and $\Delta_{\lambda}^{\alpha\beta}$ with $\alpha\neq\beta$. They characterize the formation of Cooper pairs by electrons correspondingly residing in the same and different valleys.

To figure out which type of pairing (intra- or intervalley) is more favorable in our system, we can solve the system of Gor'kov's equations,
\begin{eqnarray}
\label{Gorkov1}
\left(-\frac{\partial}{\partial \tau}-\xi_p\right)\hat{{\cal G}}(p,\tau)+\hat\Delta_{\lambda}(p)\hat F^{\dag}(p,\tau)=1,\\
\nonumber
\left(-\frac{\partial}{\partial \tau}+\xi_p\right)\hat F^{\dag}(p,\tau)+\hat\Delta^{\dag}_{\lambda}(p)
\hat{\cal{G}}(p,\tau)=0,
\end{eqnarray}
where $\xi_{p}=\pm \hbar v_0 p -\mu$ is the spectrum of electrons in doped graphene, $\mu=\hbar v_0p_F$ is the Fermi energy, $p_F=\sqrt{4\pi n_e/g_sg_v}$ is the Fermi wave vector with $n_e$ the concentration of electrons in graphene and $g_s=2$ and $g_v=2$ the spin and valley $g$-factors, and 
$\mathcal{G}^{\alpha\alpha}(p,\tau)=-\langle T_{\tau} c_{\alpha,p,\uparrow}(\tau)c^{\dag}_{\alpha,p,\uparrow}\rangle$ and $F^{\alpha\beta}(p,\tau)=\langle T_{\tau} c_{\alpha,p,\downarrow}(\tau)c_{\beta,p,\uparrow}\rangle$ are normal and anomalous Green's functions in imaginary time ($\tau=it$) representation, together with the equation for the SC gap matrix in the valley space,
\begin{eqnarray}\label{selfconsist}
\hat\Delta_{\lambda}(p)
&=&-\sum_p V_{\lambda}(p) \begin{pmatrix}
F^{11} & F^{12}\\
F^{21} & F^{22}
\end{pmatrix}(p,0)\\
\nonumber
&&-\frac{1}{2}\sum_{p,j=\pm}V_{\lambda}(p+j Q)\begin{pmatrix}
0 & F^{12}\\
F^{21} & 0
\end{pmatrix}(p,0).
\end{eqnarray}

In order to analytically estimate the critical temperatures of the intra- and intervalley pairings, let us disregard the temperature-dependent term in Eq.~(\ref{EqEffPot2b}) and solve Eqs.~(\ref{Gorkov1}) and (\ref{selfconsist}).
We can consider two limiting cases of low ($|\mu| < \omega_b$) and high ($|\mu|>\omega_b$) doping.
If we take $|\mu|\gg \omega_b$, we find
\begin{eqnarray}
\label{crtemp1}
&&T_c^\textrm{intra}=1.13~\omega_b \exp\left(-\frac{4}{V_{\lambda}^{11}(p_F)D(\mu)}\right),~~\\
&&T_c^\textrm{inter}=1.13~\omega_b \exp\left(-\frac{4}{[V_{\lambda}^{11}(p_F)+V^{12}_{\lambda}(p_F)]D(\mu)}\right),~~~~~
\label{crtemp2}
\end{eqnarray}
where $D(\mu)=(g_sg_v/2\pi\hbar^2)\mu /v_0^2$ is the density of states in graphene.
Obviously, both intra- and intervalley order parameters (and the corresponding critical temperatures) are, with good accuracy, equal to each other due to the smallness of $V^{12}_{\lambda}(p_F)= V^{21}_{\lambda}(p_F)\propto g_{Q}^2$. 
Indeed, $g_{Q}^2$ is exponentially suppressed due to large intervalley momentum $Q$. 
Thus, we can put $\Delta^{\alpha\alpha}_{\lambda}=\Delta^{\alpha\beta}_{\lambda}\equiv \Delta_\lambda/2$. Strictly speaking, one should also account for the effects of electron scattering on the non-magnetic impurities in the system. The main reason for this is that the SC correlations might be sensitive to impurities in a system with several valleys.
As shown in~\cite{RefEfetov}, intravalley pairing is usually suppressed by impurity scattering. 
In the low-doping regime, we come to the same conclusion. 

%
%

The 1b-mediated gap at zero temperature reads 
%
%
%
%
%
\begin{eqnarray}
\label{Gap1b1}
\Delta_{{1b}}^{(0)}(|\mu| > \omega_b)
&\approx&
2\omega_b \exp\left(-\frac{4}{\chi_1 D(\mu)}\right),
\\
\label{Gap1b2}
\Delta_{{1b}}^{(0)}(|\mu| < \omega_b)
&\approx& 
2|\mu| \exp\left(-\frac{4}{\chi_1 D(\mu)} +\frac{\omega_b}{|\mu|}-1\right).~~~~
\end{eqnarray}
In Eqs.~\eqref{Gap1b1} and~\eqref{Gap1b2}, we assumed $p_Fd, p_F l\ll 1$ and expanded the exponential factors in $g_p^2$. 
This assumption imposes a restriction on the maximum allowed value of $n_e$ for considered distances $d$ and $l$: ($n_e\sim {\rm min}[1/\pi d^2, 1/\pi l^2]$). 
%
%
Note that~\eqref{Gap1b1} has a standard form of the BCS gap, while the expression~\eqref{Gap1b2} is mostly determined by the doping $\mu$ rather than  $\omega_b$~\cite{kopnin}. 

The 2b-mediated gap at zero temperature is the same for both high and low doping limits,
\begin{eqnarray}
\label{Gap2b2}
\Delta_{{2b}}^{(0)}\approx 2\omega_b \exp\left(-\frac{2\pi\hbar v_0}{\chi_2} \right).~~~~
\end{eqnarray}
%
%
%
%
\begin{figure}
    \centering
    \includegraphics[width=0.48\textwidth]{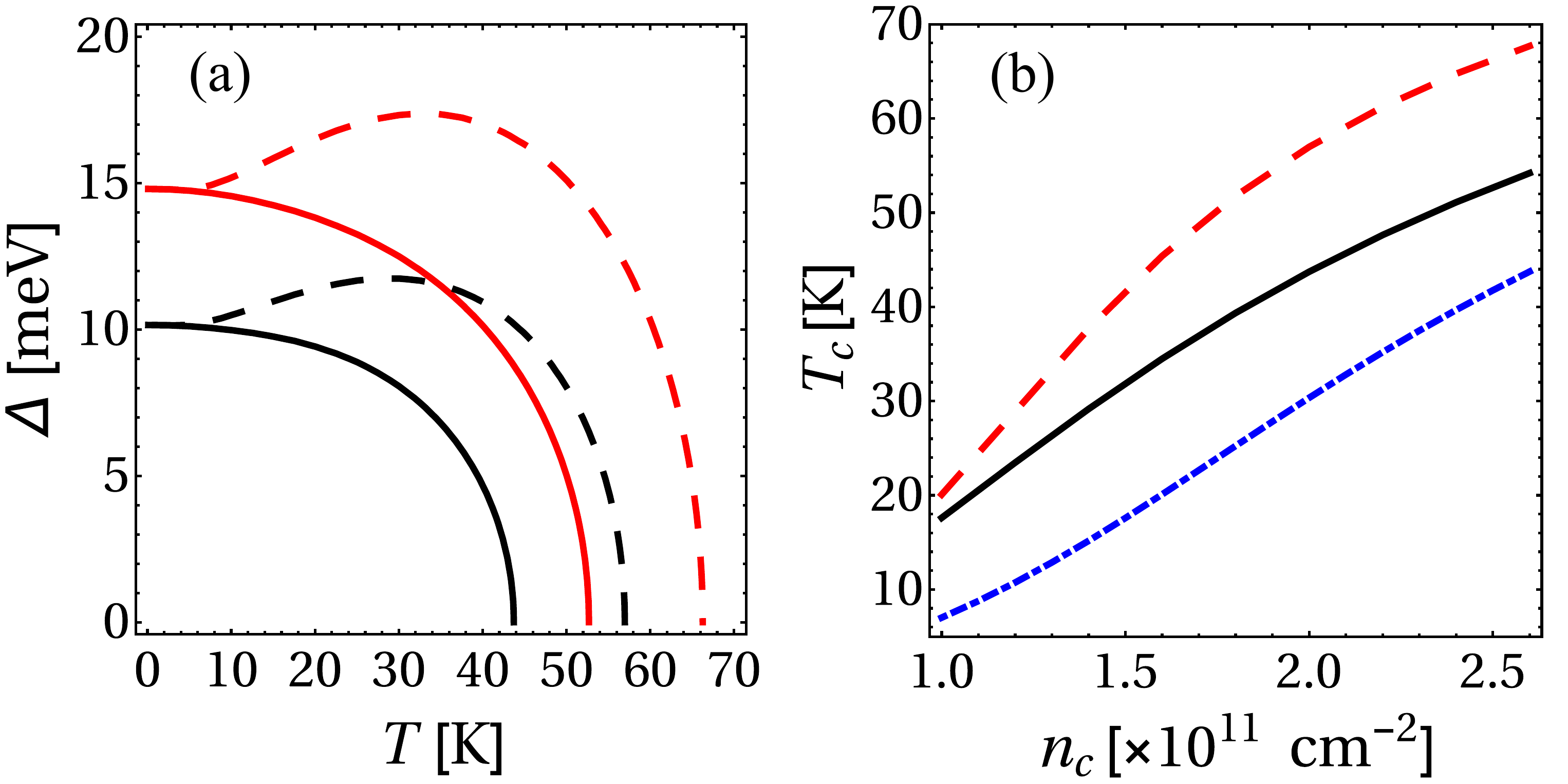}
    \caption{(a) Temperature dependence of the SC gap for bogolon-pair-mediated process with an account of the $N_q$-containing term (dashed) and without $N_q$ contribution (solid).
    We used condensate densities of $n_c=2.5\times 10^{11}$ (red) and $n_c=2.0\times 10^{11}$~cm$^{-2}$ (black),
    the density of free electrons of $n_e=7.0\times 10^{11}$~cm$^{-2}$ (thus $\mu > \omega_b$), and 
    the separation of $d=1$~nm. 
    We also accounted for the temperature dependence of $n_c$: $\Tilde{n}_c =n_c\left[1 - \left(T/T_\textrm{BEC}\right)^2\right]$ with $T_\textrm{BEC}=100$~K. 
    (b) Critical temperature of the SC transition as a function of condensate density for bogolon-pair-mediated interaction with (red dashed) and without (black solid) the $N_q$-containing contribution. {\color{black}Blue line shows the BKT transition temperature as a function of the condensate density.}
    Parameters typical for MoS$_2$ and hBN were set as follows:
    $\epsilon=4.89$, $m=0.46m_0$ (with $m_0$ the free electron mass), and $M=m_0$.
    }
    \label{Fig4}
\end{figure}
The order parameter in Eq.~(\ref{Gap2b2}) is much larger than the SC gap mediated by the 1b process. 
First, Eq.~\eqref{Gap2b2} does not contain the density of states of the Dirac electrons in graphene $D \propto |\mu|$. 
Second, the 2b-mediated SC gap 
is not determined by the chemical potential $\mu$ in the $\mu<\omega_b$ limit. 
Both of these features appear due to the nature of the 2b-mediated electron interaction, of which matrix elements are $u_{\bf p}v_{\bf p} \propto (p\xi_h)^{-1}$. 
As a result, the emerging $1/p$ term in the pairing potential [Eq.~\eqref{EqEffPot2b}] compensates the smallness due to $D(\mu)\sim p$.
It is important to note that the density of states remains small in some extended range of $k$ (not only $k\rightarrow0$), and thus 2b process is important in a wide range of wave vectors up to $\sim 10^7$~cm$^{-1}$ (or $n_e\sim10^{13}$~cm$^{-2}$).
Figure~\ref{Fig4} illustrates the full numerical solution of Eq.~\eqref{selfconsist} for the 2b process and shows that both the SC gap and the critical temperature grow with $n_c$.
%
%
%
%
As it should be, the $T_c$ we find (up to 70~K) is smaller than the possible temperature for BEC formation (around 100~K) in MoS$_2$ indirect exciton layers~\cite{Wang:2019aa}. We also note that our mean-field calculations of $T_c$ provide only the upper bound since the topological Berezinskii--Kosterlitz--Thouless (BKT) transition might occur in 2D~\cite{Berezinskii_1972, Kosterlitz_1973}. 
{\color{black}
In Fig.~\ref{Fig4}(b) we also compare the SC critical temperature with the BKT transition temperature obtained from the well-known expression (see, e.g.,~\cite{loktevreview}):  $T_{BKT}=(\pi/2)J(T_{BKT}, \Delta_{2b})$, where $J$ is a superfluid stiffness. The SC gap-dependent stiffness $J[T, \Delta_{2b}(T)]$ can be found by taking into account the small fluctuations of the phase of the order parameter~\cite{loktev}. 
}

{\color{black}
Let us also discuss how the Coulomb repulsion between electrons in graphene can affect the bogolon-mediated superconductivity. 
As it is known~\cite{mcmillan}, Coulomb repulsion leads to the renormalization of the dimensionless coupling constant in Eq.~(\ref{Gap2b2}),  $\chi_2/hv_0\rightarrow \chi_2/hv_0-D(\mu)V'_C$, where $V'_C=V_C/[1+D(\mu)V_C\log(\mu/\omega_b)]$ with $V_C$ the averaged Coulomb potential~\cite{RefEfetov} and $h=2\pi\hbar$. Taking $n_c=10^{11}\,{\rm cm}^{-2}$ and other parameters as in the caption of Fig.~\ref{Fig4}, we estimate $\chi_2/hv_0\approx 0.5$. 
In the meantime, $D(\mu)V'_C\approx 0.2$. 
Thus, we conclude that Coulomb repulsion does not destroy the 2b-mediated superconductivity in the reasonable range of parameters. 
}


{\it Conclusions.} We have studied the bogolon-mediated interaction of electrons in graphene in the vicinity of a two-dimensional Bose-condensed dipolar exciton gas. 
We developed the BCS-like bogolon-mediated electron pairing theory and calculated the critical temperature of the superconducting transition in graphene, \textcolor{black}{and compared it with the temperature of the BKT transition.} 
We showed that bogolon-pair-mediated interaction allows one to solve the problem of the smallness of the density of states in two-dimensional Dirac materials with linear spectrum at small momenta. 
\textcolor{black}{Our theory is  general and can be applied to other two-dimensional materials. 
We also expect that the  bogolon-pair-mediated process might also be combined with other mechanisms of superconductivity in 2D systems and amplify them. 
One of the application-oriented advantages of our setup is that the BEC condensate density can be used for tuning the strength of the electron pairing potential. 
It reserves an opportunity for the study of the BCS-BEC crossover in graphene.
}


{\it Acknowledgements.}
We thank J.~Rasmussen (RECON) for a critical reading of our manuscript, and E.~Savenko for the help with the figures.  
We have been partially supported by the Institute for Basic Science in Korea (Project No.~IBS-R024-D1). 
The part of this work devoted to the single-bogolon pairing was funded by the Russian Science Foundation (Project No.~17-12-01039); the work on bogolon-pair-mediated superconductivity in graphene was supported by the Russian Foundation for Basic Research (Project No.~18-29-20033).




%
%
%
%


%
%
%
%

\bibliography{library}

\begin{thebibliography}{55}%
\makeatletter
\providecommand \@ifxundefined [1]{%
 \@ifx{#1\undefined}
}%
\providecommand \@ifnum [1]{%
 \ifnum #1\expandafter \@firstoftwo
 \else \expandafter \@secondoftwo
 \fi
}%
\providecommand \@ifx [1]{%
 \ifx #1\expandafter \@firstoftwo
 \else \expandafter \@secondoftwo
 \fi
}%
\providecommand \natexlab [1]{#1}%
\providecommand \enquote  [1]{``#1''}%
\providecommand \bibnamefont  [1]{#1}%
\providecommand \bibfnamefont [1]{#1}%
\providecommand \citenamefont [1]{#1}%
\providecommand \href@noop [0]{\@secondoftwo}%
\providecommand \href [0]{\begingroup \@sanitize@url \@href}%
\providecommand \@href[1]{\@@startlink{#1}\@@href}%
\providecommand \@@href[1]{\endgroup#1\@@endlink}%
\providecommand \@sanitize@url [0]{\catcode `\\12\catcode `\$12\catcode
  `\&12\catcode `\#12\catcode `\^12\catcode `\_12\catcode `\%12\relax}%
\providecommand \@@startlink[1]{}%
\providecommand \@@endlink[0]{}%
\providecommand \url  [0]{\begingroup\@sanitize@url \@url }%
\providecommand \@url [1]{\endgroup\@href {#1}{\urlprefix }}%
\providecommand \urlprefix  [0]{URL }%
\providecommand \Eprint [0]{\href }%
\providecommand \doibase [0]{http://dx.doi.org/}%
\providecommand \selectlanguage [0]{\@gobble}%
\providecommand \bibinfo  [0]{\@secondoftwo}%
\providecommand \bibfield  [0]{\@secondoftwo}%
\providecommand \translation [1]{[#1]}%
\providecommand \BibitemOpen [0]{}%
\providecommand \bibitemStop [0]{}%
\providecommand \bibitemNoStop [0]{.\EOS\space}%
\providecommand \EOS [0]{\spacefactor3000\relax}%
\providecommand \BibitemShut  [1]{\csname bibitem#1\endcsname}%
\let\auto@bib@innerbib\@empty
\bibitem [{\citenamefont {Novoselov}\ \emph
  {et~al.}(2005{\natexlab{a}})\citenamefont {Novoselov}, \citenamefont {Jiang},
  \citenamefont {Schedin}, \citenamefont {Booth}, \citenamefont {Khotkevich},
  \citenamefont {Morozov},\ and\ \citenamefont {Geim}}]{Novoselov10451}%
  \BibitemOpen
  \bibfield  {author} {\bibinfo {author} {\bibfnamefont {K.~S.}\ \bibnamefont
  {Novoselov}}, \bibinfo {author} {\bibfnamefont {D.}~\bibnamefont {Jiang}},
  \bibinfo {author} {\bibfnamefont {F.}~\bibnamefont {Schedin}}, \bibinfo
  {author} {\bibfnamefont {T.~J.}\ \bibnamefont {Booth}}, \bibinfo {author}
  {\bibfnamefont {V.~V.}\ \bibnamefont {Khotkevich}}, \bibinfo {author}
  {\bibfnamefont {S.~V.}\ \bibnamefont {Morozov}}, \ and\ \bibinfo {author}
  {\bibfnamefont {A.~K.}\ \bibnamefont {Geim}},\ }\href {\doibase
  10.1073/pnas.0502848102} {\bibfield  {journal} {\bibinfo  {journal} {PNAS}\
  }\textbf {\bibinfo {volume} {102}},\ \bibinfo {pages} {10451} (\bibinfo
  {year} {2005}{\natexlab{a}})}\BibitemShut {NoStop}%
\bibitem [{\citenamefont {Novoselov}\ \emph {et~al.}(2004)\citenamefont
  {Novoselov}, \citenamefont {Geim}, \citenamefont {Morozov}, \citenamefont
  {Jiang}, \citenamefont {Zhang}, \citenamefont {Dubonos}, \citenamefont
  {Grigorieva},\ and\ \citenamefont {Firsov}}]{Novoselov666}%
  \BibitemOpen
  \bibfield  {author} {\bibinfo {author} {\bibfnamefont {K.~S.}\ \bibnamefont
  {Novoselov}}, \bibinfo {author} {\bibfnamefont {A.~K.}\ \bibnamefont {Geim}},
  \bibinfo {author} {\bibfnamefont {S.~V.}\ \bibnamefont {Morozov}}, \bibinfo
  {author} {\bibfnamefont {D.}~\bibnamefont {Jiang}}, \bibinfo {author}
  {\bibfnamefont {Y.}~\bibnamefont {Zhang}}, \bibinfo {author} {\bibfnamefont
  {S.~V.}\ \bibnamefont {Dubonos}}, \bibinfo {author} {\bibfnamefont {I.~V.}\
  \bibnamefont {Grigorieva}}, \ and\ \bibinfo {author} {\bibfnamefont {A.~A.}\
  \bibnamefont {Firsov}},\ }\href {\doibase 10.1126/science.1102896} {\bibfield
   {journal} {\bibinfo  {journal} {Science}\ }\textbf {\bibinfo {volume}
  {306}},\ \bibinfo {pages} {666} (\bibinfo {year} {2004})}\BibitemShut
  {NoStop}%
\bibitem [{\citenamefont {Castro~Neto}\ \emph {et~al.}(2009)\citenamefont
  {Castro~Neto}, \citenamefont {Guinea}, \citenamefont {Peres}, \citenamefont
  {Novoselov},\ and\ \citenamefont {Geim}}]{Castro-Neto:2009aa}%
  \BibitemOpen
  \bibfield  {author} {\bibinfo {author} {\bibfnamefont {A.~H.}\ \bibnamefont
  {Castro~Neto}}, \bibinfo {author} {\bibfnamefont {F.}~\bibnamefont {Guinea}},
  \bibinfo {author} {\bibfnamefont {N.~M.~R.}\ \bibnamefont {Peres}}, \bibinfo
  {author} {\bibfnamefont {K.~S.}\ \bibnamefont {Novoselov}}, \ and\ \bibinfo
  {author} {\bibfnamefont {A.~K.}\ \bibnamefont {Geim}},\ }\href {\doibase
  10.1103/RevModPhys.81.109} {\bibfield  {journal} {\bibinfo  {journal} {Rev.
  Mod. Phys.}\ }\textbf {\bibinfo {volume} {81}},\ \bibinfo {pages} {109}
  (\bibinfo {year} {2009})}\BibitemShut {NoStop}%
\bibitem [{\citenamefont {Novoselov}\ \emph
  {et~al.}(2005{\natexlab{b}})\citenamefont {Novoselov}, \citenamefont {Geim},
  \citenamefont {Morozov}, \citenamefont {Jiang}, \citenamefont {Katsnelson},
  \citenamefont {Grigorieva}, \citenamefont {V.},\ and\ \citenamefont
  {Firsov}}]{MorozovNature}%
  \BibitemOpen
  \bibfield  {author} {\bibinfo {author} {\bibfnamefont {K.~S.}\ \bibnamefont
  {Novoselov}}, \bibinfo {author} {\bibfnamefont {A.~K.}\ \bibnamefont {Geim}},
  \bibinfo {author} {\bibfnamefont {S.~V.}\ \bibnamefont {Morozov}}, \bibinfo
  {author} {\bibfnamefont {D.}~\bibnamefont {Jiang}}, \bibinfo {author}
  {\bibfnamefont {M.~I.}\ \bibnamefont {Katsnelson}}, \bibinfo {author}
  {\bibfnamefont {I.~V.}\ \bibnamefont {Grigorieva}}, \bibinfo {author}
  {\bibfnamefont {D.~S.}\ \bibnamefont {V.}}, \ and\ \bibinfo {author}
  {\bibfnamefont {A.~A.}\ \bibnamefont {Firsov}},\ }\href {\doibase
  10.1103/PhysRevB.99.115408} {\bibfield  {journal} {\bibinfo  {journal}
  {Nature}\ }\textbf {\bibinfo {volume} {438}},\ \bibinfo {pages} {197}
  (\bibinfo {year} {2005}{\natexlab{b}})}\BibitemShut {NoStop}%
\bibitem [{\citenamefont {Beenakker}(2008)}]{Beenakker2008}%
  \BibitemOpen
  \bibfield  {author} {\bibinfo {author} {\bibfnamefont {C.~W.~J.}\
  \bibnamefont {Beenakker}},\ }\href {\doibase 10.1103/RevModPhys.80.1337}
  {\bibfield  {journal} {\bibinfo  {journal} {Rev. Mod. Phys.}\ }\textbf
  {\bibinfo {volume} {80}},\ \bibinfo {pages} {1337} (\bibinfo {year}
  {2008})}\BibitemShut {NoStop}%
\bibitem [{\citenamefont {Beenakker}(2006)}]{Beenakker2006}%
  \BibitemOpen
  \bibfield  {author} {\bibinfo {author} {\bibfnamefont {C.~W.~J.}\
  \bibnamefont {Beenakker}},\ }\href {\doibase 10.1103/PhysRevLett.97.067007}
  {\bibfield  {journal} {\bibinfo  {journal} {Phys. Rev. Lett.}\ }\textbf
  {\bibinfo {volume} {97}},\ \bibinfo {pages} {067007} (\bibinfo {year}
  {2006})}\BibitemShut {NoStop}%
\bibitem [{\citenamefont {Heersche}\ \emph {et~al.}(2007)\citenamefont
  {Heersche}, \citenamefont {Jarillo-Herrero}, \citenamefont {Oostinga},
  \citenamefont {Vandersypen},\ and\ \citenamefont {Morpurgo}}]{Heersche2007}%
  \BibitemOpen
  \bibfield  {author} {\bibinfo {author} {\bibfnamefont {H.~B.}\ \bibnamefont
  {Heersche}}, \bibinfo {author} {\bibfnamefont {P.}~\bibnamefont
  {Jarillo-Herrero}}, \bibinfo {author} {\bibfnamefont {J.~B.}\ \bibnamefont
  {Oostinga}}, \bibinfo {author} {\bibfnamefont {L.~M.~K.}\ \bibnamefont
  {Vandersypen}}, \ and\ \bibinfo {author} {\bibfnamefont {A.~F.}\ \bibnamefont
  {Morpurgo}},\ }\href {\doibase 10.1038/nature05555} {\bibfield  {journal}
  {\bibinfo  {journal} {Nature}\ }\textbf {\bibinfo {volume} {446}},\ \bibinfo
  {pages} {56} (\bibinfo {year} {2007})}\BibitemShut {NoStop}%
\bibitem [{\citenamefont {Du}\ \emph {et~al.}(2008)\citenamefont {Du},
  \citenamefont {Skachko},\ and\ \citenamefont {Andrei}}]{PhysRevB.77.184507}%
  \BibitemOpen
  \bibfield  {author} {\bibinfo {author} {\bibfnamefont {X.}~\bibnamefont
  {Du}}, \bibinfo {author} {\bibfnamefont {I.}~\bibnamefont {Skachko}}, \ and\
  \bibinfo {author} {\bibfnamefont {E.~Y.}\ \bibnamefont {Andrei}},\ }\href
  {\doibase 10.1103/PhysRevB.77.184507} {\bibfield  {journal} {\bibinfo
  {journal} {Phys. Rev. B}\ }\textbf {\bibinfo {volume} {77}},\ \bibinfo
  {pages} {184507} (\bibinfo {year} {2008})}\BibitemShut {NoStop}%
\bibitem [{\citenamefont {Atienza}(2013)}]{Atienza2014}%
  \BibitemOpen
  \bibfield  {author} {\bibinfo {author} {\bibfnamefont {P.~B.}\ \bibnamefont
  {Atienza}},\ }\href@noop {} {\emph {\bibinfo {title} {Superconductivity in
  Graphene and Carbon Nanotubes: Proximity effect and nonlocal transport}}}\
  (\bibinfo  {publisher} {Springer Science \& Business Media},\ \bibinfo {year}
  {2013})\BibitemShut {NoStop}%
\bibitem [{\citenamefont {Uchoa}\ and\ \citenamefont
  {Castro~Neto}(2007)}]{CastroNetoPhysRevLett.98.146801}%
  \BibitemOpen
  \bibfield  {author} {\bibinfo {author} {\bibfnamefont {B.}~\bibnamefont
  {Uchoa}}\ and\ \bibinfo {author} {\bibfnamefont {A.~H.}\ \bibnamefont
  {Castro~Neto}},\ }\href {\doibase 10.1103/PhysRevLett.98.146801} {\bibfield
  {journal} {\bibinfo  {journal} {Phys. Rev. Lett.}\ }\textbf {\bibinfo
  {volume} {98}},\ \bibinfo {pages} {146801} (\bibinfo {year}
  {2007})}\BibitemShut {NoStop}%
\bibitem [{\citenamefont {Marchenko}\ \emph {et~al.}(2018)\citenamefont
  {Marchenko}, \citenamefont {Evtushinsky}, \citenamefont {Golias},
  \citenamefont {Varykhalov}, \citenamefont {Seyller},\ and\ \citenamefont
  {Rader}}]{Marchenko2018}%
  \BibitemOpen
  \bibfield  {author} {\bibinfo {author} {\bibfnamefont {D.}~\bibnamefont
  {Marchenko}}, \bibinfo {author} {\bibfnamefont {D.~V.}\ \bibnamefont
  {Evtushinsky}}, \bibinfo {author} {\bibfnamefont {E.}~\bibnamefont {Golias}},
  \bibinfo {author} {\bibfnamefont {A.}~\bibnamefont {Varykhalov}}, \bibinfo
  {author} {\bibfnamefont {T.}~\bibnamefont {Seyller}}, \ and\ \bibinfo
  {author} {\bibfnamefont {O.}~\bibnamefont {Rader}},\ }\href {\doibase
  10.1126/sciadv.aau0059} {\bibfield  {journal} {\bibinfo  {journal} {Sci.
  Advances}\ }\textbf {\bibinfo {volume} {4}} (\bibinfo {year} {2018}),\
  10.1126/sciadv.aau0059}\BibitemShut {NoStop}%
\bibitem [{\citenamefont {Laussy}\ \emph {et~al.}(2010)\citenamefont {Laussy},
  \citenamefont {Kavokin},\ and\ \citenamefont {Shelykh}}]{Laussy:2010aa}%
  \BibitemOpen
  \bibfield  {author} {\bibinfo {author} {\bibfnamefont {F.~P.}\ \bibnamefont
  {Laussy}}, \bibinfo {author} {\bibfnamefont {A.~V.}\ \bibnamefont {Kavokin}},
  \ and\ \bibinfo {author} {\bibfnamefont {I.~A.}\ \bibnamefont {Shelykh}},\
  }\href {\doibase 10.1103/PhysRevLett.104.106402} {\bibfield  {journal}
  {\bibinfo  {journal} {Phys. Rev. Lett.}\ }\textbf {\bibinfo {volume} {104}},\
  \bibinfo {pages} {106402} (\bibinfo {year} {2010})}\BibitemShut {NoStop}%
\bibitem [{\citenamefont {Cotle{\c t}}\ \emph {et~al.}(2016)\citenamefont
  {Cotle{\c t}}, \citenamefont {Zeytino\ifmmode~\check{g}\else \v{g}\fi{}lu},
  \citenamefont {Sigrist}, \citenamefont {Demler},\ and\ \citenamefont
  {Imamo\ifmmode~\check{g}\else \v{g}\fi{}lu}}]{Cotleifmmode-telse-tfi:2016aa}%
  \BibitemOpen
  \bibfield  {author} {\bibinfo {author} {\bibfnamefont {O.}~\bibnamefont
  {Cotle{\c t}}}, \bibinfo {author} {\bibfnamefont {S.}~\bibnamefont
  {Zeytino\ifmmode~\check{g}\else \v{g}\fi{}lu}}, \bibinfo {author}
  {\bibfnamefont {M.}~\bibnamefont {Sigrist}}, \bibinfo {author} {\bibfnamefont
  {E.}~\bibnamefont {Demler}}, \ and\ \bibinfo {author} {\bibfnamefont
  {A.}~\bibnamefont {Imamo\ifmmode~\check{g}\else \v{g}\fi{}lu}},\ }\href
  {\doibase 10.1103/PhysRevB.93.054510} {\bibfield  {journal} {\bibinfo
  {journal} {Phys. Rev. B}\ }\textbf {\bibinfo {volume} {93}},\ \bibinfo
  {pages} {054510} (\bibinfo {year} {2016})}\BibitemShut {NoStop}%
\bibitem [{\citenamefont {Villegas}\ \emph {et~al.}(2019)\citenamefont
  {Villegas}, \citenamefont {Sun}, \citenamefont {Kovalev},\ and\ \citenamefont
  {Savenko}}]{OurPRLBG}%
  \BibitemOpen
  \bibfield  {author} {\bibinfo {author} {\bibfnamefont {K.~H.~A.}\
  \bibnamefont {Villegas}}, \bibinfo {author} {\bibfnamefont {M.}~\bibnamefont
  {Sun}}, \bibinfo {author} {\bibfnamefont {V.~M.}\ \bibnamefont {Kovalev}}, \
  and\ \bibinfo {author} {\bibfnamefont {I.~G.}\ \bibnamefont {Savenko}},\
  }\href {\doibase 10.1103/PhysRevLett.123.095301} {\bibfield  {journal}
  {\bibinfo  {journal} {Phys. Rev. Lett.}\ }\textbf {\bibinfo {volume} {123}},\
  \bibinfo {pages} {095301} (\bibinfo {year} {2019})}\BibitemShut {NoStop}%
\bibitem [{\citenamefont {Skopelitis}\ \emph {et~al.}(2018)\citenamefont
  {Skopelitis}, \citenamefont {Cherotchenko}, \citenamefont {Kavokin},\ and\
  \citenamefont {Posazhennikova}}]{Skopelitis:2018aa}%
  \BibitemOpen
  \bibfield  {author} {\bibinfo {author} {\bibfnamefont {P.}~\bibnamefont
  {Skopelitis}}, \bibinfo {author} {\bibfnamefont {E.~D.}\ \bibnamefont
  {Cherotchenko}}, \bibinfo {author} {\bibfnamefont {A.~V.}\ \bibnamefont
  {Kavokin}}, \ and\ \bibinfo {author} {\bibfnamefont {A.}~\bibnamefont
  {Posazhennikova}},\ }\href {\doibase 10.1103/PhysRevLett.120.107001}
  {\bibfield  {journal} {\bibinfo  {journal} {Phys. Rev. Lett.}\ }\textbf
  {\bibinfo {volume} {120}},\ \bibinfo {pages} {107001} (\bibinfo {year}
  {2018})}\BibitemShut {NoStop}%
\bibitem [{\citenamefont {Mahan}(1990)}]{Mahan}%
  \BibitemOpen
  \bibfield  {author} {\bibinfo {author} {\bibfnamefont {G.~D.}\ \bibnamefont
  {Mahan}},\ }\href@noop {} {\emph {\bibinfo {title} {{Many-Particle
  Physics}}}}\ (\bibinfo  {publisher} {Plenum Press. New York and London},\
  \bibinfo {year} {1990})\BibitemShut {NoStop}%
\bibitem [{\citenamefont {Cao}\ \emph {et~al.}(2018)\citenamefont {Cao},
  \citenamefont {Fatemi}, \citenamefont {Fang}, \citenamefont {Watanabe},
  \citenamefont {Taniguchi}, \citenamefont {Kaxiras},\ and\ \citenamefont
  {Jarillo-Herrero}}]{Caoa2018}%
  \BibitemOpen
  \bibfield  {author} {\bibinfo {author} {\bibfnamefont {Y.}~\bibnamefont
  {Cao}}, \bibinfo {author} {\bibfnamefont {V.}~\bibnamefont {Fatemi}},
  \bibinfo {author} {\bibfnamefont {S.}~\bibnamefont {Fang}}, \bibinfo {author}
  {\bibfnamefont {K.}~\bibnamefont {Watanabe}}, \bibinfo {author}
  {\bibfnamefont {T.}~\bibnamefont {Taniguchi}}, \bibinfo {author}
  {\bibfnamefont {E.}~\bibnamefont {Kaxiras}}, \ and\ \bibinfo {author}
  {\bibfnamefont {P.}~\bibnamefont {Jarillo-Herrero}},\ }\href {\doibase
  10.1038/nature26160} {\bibfield  {journal} {\bibinfo  {journal} {Nature
  (London)}\ }\textbf {\bibinfo {volume} {556}},\ \bibinfo {pages} {43}
  (\bibinfo {year} {2018})}\BibitemShut {NoStop}%
\bibitem [{\citenamefont {Feldman}(2019)}]{Yankowitz2019}%
  \BibitemOpen
  \bibfield  {author} {\bibinfo {author} {\bibfnamefont {B.~E.}\ \bibnamefont
  {Feldman}},\ }\href {\doibase 10.1126/science.aaw4642} {\bibfield  {journal}
  {\bibinfo  {journal} {Science}\ }\textbf {\bibinfo {volume} {363}},\ \bibinfo
  {pages} {1035} (\bibinfo {year} {2019})}\BibitemShut {NoStop}%
\bibitem [{\citenamefont {Chen}\ \emph {et~al.}(2019)\citenamefont {Chen},
  \citenamefont {Sharpe}, \citenamefont {Gallagher}, \citenamefont {Rosen},
  \citenamefont {Fox}, \citenamefont {Jiang}, \citenamefont {Lyu},
  \citenamefont {Li}, \citenamefont {Watanabe}, \citenamefont {Taniguchi},
  \citenamefont {Jung}, \citenamefont {Shi}, \citenamefont {Goldhaber-Gordon},
  \citenamefont {Zhang},\ and\ \citenamefont {Wang}}]{Chen2019}%
  \BibitemOpen
  \bibfield  {author} {\bibinfo {author} {\bibfnamefont {G.}~\bibnamefont
  {Chen}}, \bibinfo {author} {\bibfnamefont {A.~L.}\ \bibnamefont {Sharpe}},
  \bibinfo {author} {\bibfnamefont {P.}~\bibnamefont {Gallagher}}, \bibinfo
  {author} {\bibfnamefont {I.~T.}\ \bibnamefont {Rosen}}, \bibinfo {author}
  {\bibfnamefont {E.~J.}\ \bibnamefont {Fox}}, \bibinfo {author} {\bibfnamefont
  {L.}~\bibnamefont {Jiang}}, \bibinfo {author} {\bibfnamefont
  {B.}~\bibnamefont {Lyu}}, \bibinfo {author} {\bibfnamefont {H.}~\bibnamefont
  {Li}}, \bibinfo {author} {\bibfnamefont {K.}~\bibnamefont {Watanabe}},
  \bibinfo {author} {\bibfnamefont {T.}~\bibnamefont {Taniguchi}}, \bibinfo
  {author} {\bibfnamefont {J.}~\bibnamefont {Jung}}, \bibinfo {author}
  {\bibfnamefont {Z.}~\bibnamefont {Shi}}, \bibinfo {author} {\bibfnamefont
  {D.}~\bibnamefont {Goldhaber-Gordon}}, \bibinfo {author} {\bibfnamefont
  {Y.}~\bibnamefont {Zhang}}, \ and\ \bibinfo {author} {\bibfnamefont
  {F.}~\bibnamefont {Wang}},\ }\href {\doibase 10.1038/nature26160} {\bibfield
  {journal} {\bibinfo  {journal} {Nature (London)}\ }\textbf {\bibinfo {volume}
  {572}},\ \bibinfo {pages} {215} (\bibinfo {year} {2019})}\BibitemShut
  {NoStop}%
\bibitem [{\citenamefont {Kopnin}\ \emph {et~al.}(2011)\citenamefont {Kopnin},
  \citenamefont {Heikkil\"a},\ and\ \citenamefont {Volovik}}]{Kopnin2011}%
  \BibitemOpen
  \bibfield  {author} {\bibinfo {author} {\bibfnamefont {N.~B.}\ \bibnamefont
  {Kopnin}}, \bibinfo {author} {\bibfnamefont {T.~T.}\ \bibnamefont
  {Heikkil\"a}}, \ and\ \bibinfo {author} {\bibfnamefont {G.~E.}\ \bibnamefont
  {Volovik}},\ }\href {\doibase 10.1103/PhysRevB.83.220503} {\bibfield
  {journal} {\bibinfo  {journal} {Phys. Rev. B}\ }\textbf {\bibinfo {volume}
  {83}},\ \bibinfo {pages} {220503} (\bibinfo {year} {2011})}\BibitemShut
  {NoStop}%
\bibitem [{\citenamefont {Mu\~noz}\ \emph {et~al.}(2013)\citenamefont
  {Mu\~noz}, \citenamefont {Covaci},\ and\ \citenamefont
  {Peeters}}]{Munoz2013}%
  \BibitemOpen
  \bibfield  {author} {\bibinfo {author} {\bibfnamefont {W.~A.}\ \bibnamefont
  {Mu\~noz}}, \bibinfo {author} {\bibfnamefont {L.}~\bibnamefont {Covaci}}, \
  and\ \bibinfo {author} {\bibfnamefont {F.~M.}\ \bibnamefont {Peeters}},\
  }\href {\doibase 10.1103/PhysRevB.87.134509} {\bibfield  {journal} {\bibinfo
  {journal} {Phys. Rev. B}\ }\textbf {\bibinfo {volume} {87}},\ \bibinfo
  {pages} {134509} (\bibinfo {year} {2013})}\BibitemShut {NoStop}%
\bibitem [{\citenamefont {Bistritzer}\ and\ \citenamefont
  {MacDonald}(2011)}]{Bistritzer2011}%
  \BibitemOpen
  \bibfield  {author} {\bibinfo {author} {\bibfnamefont {R.}~\bibnamefont
  {Bistritzer}}\ and\ \bibinfo {author} {\bibfnamefont {A.~H.}\ \bibnamefont
  {MacDonald}},\ }\href {\doibase 10.1073/pnas.1108174108} {\bibfield
  {journal} {\bibinfo  {journal} {PNAS}\ }\textbf {\bibinfo {volume} {108}},\
  \bibinfo {pages} {12233} (\bibinfo {year} {2011})}\BibitemShut {NoStop}%
\bibitem [{\citenamefont {Wang}\ \emph {et~al.}(2020)\citenamefont {Wang},
  \citenamefont {Zheng}, \citenamefont {Chen}, \citenamefont {Huang},
  \citenamefont {Kartashov}, \citenamefont {Torner}, \citenamefont {Konotop},\
  and\ \citenamefont {Ye}}]{WangNature2020}%
  \BibitemOpen
  \bibfield  {author} {\bibinfo {author} {\bibfnamefont {P.}~\bibnamefont
  {Wang}}, \bibinfo {author} {\bibfnamefont {Y.}~\bibnamefont {Zheng}},
  \bibinfo {author} {\bibfnamefont {X.}~\bibnamefont {Chen}}, \bibinfo {author}
  {\bibfnamefont {C.}~\bibnamefont {Huang}}, \bibinfo {author} {\bibfnamefont
  {Y.~V.}\ \bibnamefont {Kartashov}}, \bibinfo {author} {\bibfnamefont
  {L.}~\bibnamefont {Torner}}, \bibinfo {author} {\bibfnamefont {V.~V.}\
  \bibnamefont {Konotop}}, \ and\ \bibinfo {author} {\bibfnamefont
  {F.}~\bibnamefont {Ye}},\ }\href {\doibase 10.1038/s41586-019-1851-6}
  {\bibfield  {journal} {\bibinfo  {journal} {Nature (London)}\ }\textbf
  {\bibinfo {volume} {577}},\ \bibinfo {pages} {42} (\bibinfo {year}
  {2020})}\BibitemShut {NoStop}%
\bibitem [{\citenamefont {Lopes~dos Santos}\ \emph {et~al.}(2012)\citenamefont
  {Lopes~dos Santos}, \citenamefont {Peres},\ and\ \citenamefont
  {Castro~Neto}}]{PhysRevB.86.155449}%
  \BibitemOpen
  \bibfield  {author} {\bibinfo {author} {\bibfnamefont {J.~M.~B.}\
  \bibnamefont {Lopes~dos Santos}}, \bibinfo {author} {\bibfnamefont
  {N.~M.~R.}\ \bibnamefont {Peres}}, \ and\ \bibinfo {author} {\bibfnamefont
  {A.~H.}\ \bibnamefont {Castro~Neto}},\ }\href {\doibase
  10.1103/PhysRevB.86.155449} {\bibfield  {journal} {\bibinfo  {journal} {Phys.
  Rev. B}\ }\textbf {\bibinfo {volume} {86}},\ \bibinfo {pages} {155449}
  (\bibinfo {year} {2012})}\BibitemShut {NoStop}%
\bibitem [{\citenamefont {Wen-Hao}\ \emph {et~al.}(2014)\citenamefont
  {Wen-Hao}, \citenamefont {Yi}, \citenamefont {Jin-Song}, \citenamefont
  {Fang-Sen}, \citenamefont {Ming-Hua}, \citenamefont {Yan-Fei}, \citenamefont
  {Hui-Min}, \citenamefont {Jun-Ping}, \citenamefont {Ying}, \citenamefont
  {Hui-Chao}, \citenamefont {Takeshi}, \citenamefont {Akihiko}, \citenamefont
  {Zhi}, \citenamefont {Hao}, \citenamefont {Chen-Jia}, \citenamefont {Meng},
  \citenamefont {Qing-Yan}, \citenamefont {Ke}, \citenamefont {Shuai-Hua},
  \citenamefont {Xi}, \citenamefont {Jun-Feng}, \citenamefont {Zheng-Cai},
  \citenamefont {Liang}, \citenamefont {Ya-Yu}, \citenamefont {Jian},
  \citenamefont {Li-Li}, \citenamefont {Ming-Wei}, \citenamefont {Qi-Kun},\
  and\ \citenamefont {Xu-Cun}}]{ZHANG17401}%
  \BibitemOpen
  \bibfield  {author} {\bibinfo {author} {\bibfnamefont {Z.}~\bibnamefont
  {Wen-Hao}}, \bibinfo {author} {\bibfnamefont {S.}~\bibnamefont {Yi}},
  \bibinfo {author} {\bibfnamefont {Z.}~\bibnamefont {Jin-Song}}, \bibinfo
  {author} {\bibfnamefont {L.}~\bibnamefont {Fang-Sen}}, \bibinfo {author}
  {\bibfnamefont {G.}~\bibnamefont {Ming-Hua}}, \bibinfo {author}
  {\bibfnamefont {Z.}~\bibnamefont {Yan-Fei}}, \bibinfo {author} {\bibfnamefont
  {Z.}~\bibnamefont {Hui-Min}}, \bibinfo {author} {\bibfnamefont
  {P.}~\bibnamefont {Jun-Ping}}, \bibinfo {author} {\bibfnamefont
  {X.}~\bibnamefont {Ying}}, \bibinfo {author} {\bibfnamefont {W.}~\bibnamefont
  {Hui-Chao}}, \bibinfo {author} {\bibfnamefont {F.}~\bibnamefont {Takeshi}},
  \bibinfo {author} {\bibfnamefont {H.}~\bibnamefont {Akihiko}}, \bibinfo
  {author} {\bibfnamefont {L.}~\bibnamefont {Zhi}}, \bibinfo {author}
  {\bibfnamefont {D.}~\bibnamefont {Hao}}, \bibinfo {author} {\bibfnamefont
  {T.}~\bibnamefont {Chen-Jia}}, \bibinfo {author} {\bibfnamefont
  {W.}~\bibnamefont {Meng}}, \bibinfo {author} {\bibfnamefont {W.}~\bibnamefont
  {Qing-Yan}}, \bibinfo {author} {\bibfnamefont {H.}~\bibnamefont {Ke}},
  \bibinfo {author} {\bibfnamefont {J.}~\bibnamefont {Shuai-Hua}}, \bibinfo
  {author} {\bibfnamefont {C.}~\bibnamefont {Xi}}, \bibinfo {author}
  {\bibfnamefont {W.}~\bibnamefont {Jun-Feng}}, \bibinfo {author}
  {\bibfnamefont {X.}~\bibnamefont {Zheng-Cai}}, \bibinfo {author}
  {\bibfnamefont {L.}~\bibnamefont {Liang}}, \bibinfo {author} {\bibfnamefont
  {W.}~\bibnamefont {Ya-Yu}}, \bibinfo {author} {\bibfnamefont
  {W.}~\bibnamefont {Jian}}, \bibinfo {author} {\bibfnamefont {W.}~\bibnamefont
  {Li-Li}}, \bibinfo {author} {\bibfnamefont {C.}~\bibnamefont {Ming-Wei}},
  \bibinfo {author} {\bibfnamefont {X.}~\bibnamefont {Qi-Kun}}, \ and\ \bibinfo
  {author} {\bibfnamefont {M.}~\bibnamefont {Xu-Cun}},\ }\href {\doibase
  10.1088/0256-307X/31/1/017401} {\bibfield  {journal} {\bibinfo  {journal}
  {Chinese Physics Letters}\ }\textbf {\bibinfo {volume} {31}},\ \bibinfo {eid}
  {017401} (\bibinfo {year} {2014})}\BibitemShut {NoStop}%
\bibitem [{\citenamefont {Honerkamp}(2008)}]{exotic1}%
  \BibitemOpen
  \bibfield  {author} {\bibinfo {author} {\bibfnamefont {C.}~\bibnamefont
  {Honerkamp}},\ }\href {\doibase 10.1103/PhysRevLett.100.146404} {\bibfield
  {journal} {\bibinfo  {journal} {Phys. Rev. Lett.}\ }\textbf {\bibinfo
  {volume} {100}},\ \bibinfo {pages} {146404} (\bibinfo {year}
  {2008})}\BibitemShut {NoStop}%
\bibitem [{\citenamefont {Nandkishore}\ \emph {et~al.}(2012)\citenamefont
  {Nandkishore}, \citenamefont {Levitov},\ and\ \citenamefont
  {Chubukov}}]{exotic2}%
  \BibitemOpen
  \bibfield  {author} {\bibinfo {author} {\bibfnamefont {R.}~\bibnamefont
  {Nandkishore}}, \bibinfo {author} {\bibfnamefont {L.~S.}\ \bibnamefont
  {Levitov}}, \ and\ \bibinfo {author} {\bibfnamefont {A.~V.}\ \bibnamefont
  {Chubukov}},\ }\href {\doibase 10.1038/nphys2208} {\bibfield  {journal}
  {\bibinfo  {journal} {Nature Phys.}\ }\textbf {\bibinfo {volume} {8}},\
  \bibinfo {pages} {158} (\bibinfo {year} {2012})}\BibitemShut {NoStop}%
\bibitem [{\citenamefont {Roy}\ and\ \citenamefont {Juri\ifmmode \check{c}\else
  \v{c}\fi{}i\ifmmode~\acute{c}\else \'{c}\fi{}}(2014)}]{exotic3}%
  \BibitemOpen
  \bibfield  {author} {\bibinfo {author} {\bibfnamefont {B.}~\bibnamefont
  {Roy}}\ and\ \bibinfo {author} {\bibfnamefont {V.}~\bibnamefont {Juri\ifmmode
  \check{c}\else \v{c}\fi{}i\ifmmode~\acute{c}\else \'{c}\fi{}}},\ }\href
  {\doibase 10.1103/PhysRevB.90.041413} {\bibfield  {journal} {\bibinfo
  {journal} {Phys. Rev. B}\ }\textbf {\bibinfo {volume} {90}},\ \bibinfo
  {pages} {041413} (\bibinfo {year} {2014})}\BibitemShut {NoStop}%
\bibitem [{\citenamefont {Einenkel}\ and\ \citenamefont
  {Efetov}(2011)}]{RefEfetov}%
  \BibitemOpen
  \bibfield  {author} {\bibinfo {author} {\bibfnamefont {M.}~\bibnamefont
  {Einenkel}}\ and\ \bibinfo {author} {\bibfnamefont {K.~B.}\ \bibnamefont
  {Efetov}},\ }\href {\doibase 10.1103/PhysRevB.84.214508} {\bibfield
  {journal} {\bibinfo  {journal} {Phys. Rev. B}\ }\textbf {\bibinfo {volume}
  {84}},\ \bibinfo {pages} {214508} (\bibinfo {year} {2011})}\BibitemShut
  {NoStop}%
\bibitem [{\citenamefont {Das~Sarma}\ \emph {et~al.}(2011)\citenamefont
  {Das~Sarma}, \citenamefont {Adam}, \citenamefont {Hwang},\ and\ \citenamefont
  {Rossi}}]{Das-Sarma:2011aa}%
  \BibitemOpen
  \bibfield  {author} {\bibinfo {author} {\bibfnamefont {S.}~\bibnamefont
  {Das~Sarma}}, \bibinfo {author} {\bibfnamefont {S.}~\bibnamefont {Adam}},
  \bibinfo {author} {\bibfnamefont {E.~H.}\ \bibnamefont {Hwang}}, \ and\
  \bibinfo {author} {\bibfnamefont {E.}~\bibnamefont {Rossi}},\ }\href
  {\doibase 10.1103/RevModPhys.83.407} {\bibfield  {journal} {\bibinfo
  {journal} {Rev. Mod. Phys.}\ }\textbf {\bibinfo {volume} {83}},\ \bibinfo
  {pages} {407} (\bibinfo {year} {2011})}\BibitemShut {NoStop}%
\bibitem [{\citenamefont {Butov}(2017)}]{Butov2017}%
  \BibitemOpen
  \bibfield  {author} {\bibinfo {author} {\bibfnamefont {L.~V.}\ \bibnamefont
  {Butov}},\ }\href {\doibase 10.1016/j.spmi.2016.12.035} {\bibfield  {journal}
  {\bibinfo  {journal} {Superlatt. and Microstr.}\ }\textbf {\bibinfo {volume}
  {108}},\ \bibinfo {pages} {2} (\bibinfo {year} {2017})}\BibitemShut {NoStop}%
\bibitem [{\citenamefont {Fogler}\ \emph {et~al.}(2016)\citenamefont {Fogler},
  \citenamefont {Butov},\ and\ \citenamefont {Novoselov}}]{Fogler2014}%
  \BibitemOpen
  \bibfield  {author} {\bibinfo {author} {\bibfnamefont {M.~M.}\ \bibnamefont
  {Fogler}}, \bibinfo {author} {\bibfnamefont {L.}~\bibnamefont {Butov}}, \
  and\ \bibinfo {author} {\bibfnamefont {K.~S.}\ \bibnamefont {Novoselov}},\
  }\href {\doibase 10.1038/ncomms5555} {\bibfield  {journal} {\bibinfo
  {journal} {Nature Comm.}\ }\textbf {\bibinfo {volume} {5}},\ \bibinfo {pages}
  {4555} (\bibinfo {year} {2016})}\BibitemShut {NoStop}%
\bibitem [{\citenamefont {Berman}\ and\ \citenamefont
  {Kezerashvili}(2016)}]{Berman2016}%
  \BibitemOpen
  \bibfield  {author} {\bibinfo {author} {\bibfnamefont {O.~L.}\ \bibnamefont
  {Berman}}\ and\ \bibinfo {author} {\bibfnamefont {R.~Y.}\ \bibnamefont
  {Kezerashvili}},\ }\href {\doibase 10.1103/PhysRevB.93.245410} {\bibfield
  {journal} {\bibinfo  {journal} {Phys. Rev. B}\ }\textbf {\bibinfo {volume}
  {93}},\ \bibinfo {pages} {245410} (\bibinfo {year} {2016})}\BibitemShut
  {NoStop}%
\bibitem [{\citenamefont {Kasprzak}\ \emph {et~al.}(2006)\citenamefont
  {Kasprzak}, \citenamefont {Richard}, \citenamefont {Kundermann},
  \citenamefont {Baas}, \citenamefont {Jeambrun}, \citenamefont {Keeling},
  \citenamefont {Marchetti}, \citenamefont {Szyma{\'n}ska}, \citenamefont
  {Andr{\'e}}, \citenamefont {Staehli}, \citenamefont {Savona}, \citenamefont
  {Littlewood}, \citenamefont {Deveaud},\ and\ \citenamefont
  {Dang}}]{Kasprzak:2006aa}%
  \BibitemOpen
  \bibfield  {author} {\bibinfo {author} {\bibfnamefont {J.}~\bibnamefont
  {Kasprzak}}, \bibinfo {author} {\bibfnamefont {M.}~\bibnamefont {Richard}},
  \bibinfo {author} {\bibfnamefont {S.}~\bibnamefont {Kundermann}}, \bibinfo
  {author} {\bibfnamefont {A.}~\bibnamefont {Baas}}, \bibinfo {author}
  {\bibfnamefont {P.}~\bibnamefont {Jeambrun}}, \bibinfo {author}
  {\bibfnamefont {J.~M.~J.}\ \bibnamefont {Keeling}}, \bibinfo {author}
  {\bibfnamefont {F.~M.}\ \bibnamefont {Marchetti}}, \bibinfo {author}
  {\bibfnamefont {M.~H.}\ \bibnamefont {Szyma{\'n}ska}}, \bibinfo {author}
  {\bibfnamefont {R.}~\bibnamefont {Andr{\'e}}}, \bibinfo {author}
  {\bibfnamefont {J.~L.}\ \bibnamefont {Staehli}}, \bibinfo {author}
  {\bibfnamefont {V.}~\bibnamefont {Savona}}, \bibinfo {author} {\bibfnamefont
  {P.~B.}\ \bibnamefont {Littlewood}}, \bibinfo {author} {\bibfnamefont
  {B.}~\bibnamefont {Deveaud}}, \ and\ \bibinfo {author} {\bibfnamefont
  {L.~S.}\ \bibnamefont {Dang}},\ }\href
  {http://dx.doi.org/10.1038/nature05131} {\bibfield  {journal} {\bibinfo
  {journal} {Nature (London)}\ }\textbf {\bibinfo {volume} {443}},\ \bibinfo
  {pages} {409 EP } (\bibinfo {year} {2006})}\BibitemShut {NoStop}%
\bibitem [{C1()}]{C1}%
  \BibitemOpen
  \href@noop {} {\ }\BibitemShut {NoStop}%
\bibitem [{\citenamefont {Boev}\ \emph {et~al.}(2016)\citenamefont {Boev},
  \citenamefont {Kovalev},\ and\ \citenamefont {Savenko}}]{Boev:2016aa}%
  \BibitemOpen
  \bibfield  {author} {\bibinfo {author} {\bibfnamefont {M.~V.}\ \bibnamefont
  {Boev}}, \bibinfo {author} {\bibfnamefont {V.~M.}\ \bibnamefont {Kovalev}}, \
  and\ \bibinfo {author} {\bibfnamefont {I.~G.}\ \bibnamefont {Savenko}},\
  }\href {\doibase 10.1103/PhysRevB.94.241408} {\bibfield  {journal} {\bibinfo
  {journal} {Phys. Rev. B}\ }\textbf {\bibinfo {volume} {94}},\ \bibinfo
  {pages} {241408} (\bibinfo {year} {2016})}\BibitemShut {NoStop}%
\bibitem [{\citenamefont {Matuszewski}\ \emph {et~al.}(2012)\citenamefont
  {Matuszewski}, \citenamefont {Taylor},\ and\ \citenamefont
  {Kavokin}}]{Matuszewski:2012aa}%
  \BibitemOpen
  \bibfield  {author} {\bibinfo {author} {\bibfnamefont {M.}~\bibnamefont
  {Matuszewski}}, \bibinfo {author} {\bibfnamefont {T.}~\bibnamefont {Taylor}},
  \ and\ \bibinfo {author} {\bibfnamefont {A.~V.}\ \bibnamefont {Kavokin}},\
  }\href {\doibase 10.1103/PhysRevLett.108.060401} {\bibfield  {journal}
  {\bibinfo  {journal} {Phys. Rev. Lett.}\ }\textbf {\bibinfo {volume} {108}},\
  \bibinfo {pages} {060401} (\bibinfo {year} {2012})}\BibitemShut {NoStop}%
\bibitem [{\citenamefont {Sun}\ \emph {et~al.}(2019)\citenamefont {Sun},
  \citenamefont {Villegas}, \citenamefont {Kovalev},\ and\ \citenamefont
  {Savenko}}]{Sun2018}%
  \BibitemOpen
  \bibfield  {author} {\bibinfo {author} {\bibfnamefont {M.}~\bibnamefont
  {Sun}}, \bibinfo {author} {\bibfnamefont {K.~H.~A.}\ \bibnamefont
  {Villegas}}, \bibinfo {author} {\bibfnamefont {V.~M.}\ \bibnamefont
  {Kovalev}}, \ and\ \bibinfo {author} {\bibfnamefont {I.~G.}\ \bibnamefont
  {Savenko}},\ }\href {\doibase 10.1103/PhysRevB.99.115408} {\bibfield
  {journal} {\bibinfo  {journal} {Phys. Rev. B}\ }\textbf {\bibinfo {volume}
  {99}},\ \bibinfo {pages} {115408} (\bibinfo {year} {2019})}\BibitemShut
  {NoStop}%
\bibitem [{C6()}]{C6}%
  \BibitemOpen
  \href@noop {} {\ }\BibitemShut {NoStop}%
\bibitem [{\citenamefont {Giorgini}(1998)}]{Giorgini:1998aa}%
  \BibitemOpen
  \bibfield  {author} {\bibinfo {author} {\bibfnamefont {S.}~\bibnamefont
  {Giorgini}},\ }\href {\doibase 10.1103/PhysRevA.57.2949} {\bibfield
  {journal} {\bibinfo  {journal} {Phys. Rev. A}\ }\textbf {\bibinfo {volume}
  {57}},\ \bibinfo {pages} {2949} (\bibinfo {year} {1998})}\BibitemShut
  {NoStop}%
\bibitem [{SMB()}]{SMBG}%
  \BibitemOpen
  \href@noop {} {\ }\BibitemShut {NoStop}%
\bibitem [{C2()}]{C2}%
  \BibitemOpen
  \href@noop {} {\ }\BibitemShut {NoStop}%
\bibitem [{\citenamefont {Kopnin}\ and\ \citenamefont {Sonin}(2008)}]{kopnin}%
  \BibitemOpen
  \bibfield  {author} {\bibinfo {author} {\bibfnamefont {N.~B.}\ \bibnamefont
  {Kopnin}}\ and\ \bibinfo {author} {\bibfnamefont {E.~B.}\ \bibnamefont
  {Sonin}},\ }\href {\doibase 10.1103/PhysRevLett.100.246808} {\bibfield
  {journal} {\bibinfo  {journal} {Phys. Rev. Lett.}\ }\textbf {\bibinfo
  {volume} {100}},\ \bibinfo {pages} {246808} (\bibinfo {year}
  {2008})}\BibitemShut {NoStop}%
\bibitem [{\citenamefont {Wang}\ \emph
  {et~al.}(2019{\natexlab{a}})\citenamefont {Wang}, \citenamefont {Rhodes},
  \citenamefont {Watanabe}, \citenamefont {Taniguchi}, \citenamefont {Hone},
  \citenamefont {Shan},\ and\ \citenamefont {Mak}}]{Wang:2019aa}%
  \BibitemOpen
  \bibfield  {author} {\bibinfo {author} {\bibfnamefont {Z.}~\bibnamefont
  {Wang}}, \bibinfo {author} {\bibfnamefont {D.~A.}\ \bibnamefont {Rhodes}},
  \bibinfo {author} {\bibfnamefont {K.}~\bibnamefont {Watanabe}}, \bibinfo
  {author} {\bibfnamefont {T.}~\bibnamefont {Taniguchi}}, \bibinfo {author}
  {\bibfnamefont {J.~C.}\ \bibnamefont {Hone}}, \bibinfo {author}
  {\bibfnamefont {J.}~\bibnamefont {Shan}}, \ and\ \bibinfo {author}
  {\bibfnamefont {K.~F.}\ \bibnamefont {Mak}},\ }\href {\doibase
  https://doi.org/10.1038/s41586-019-1591-7} {\bibfield  {journal} {\bibinfo
  {journal} {Nature}\ }\textbf {\bibinfo {volume} {574}},\ \bibinfo {pages}
  {76} (\bibinfo {year} {2019}{\natexlab{a}})}\BibitemShut {NoStop}%
\bibitem [{\citenamefont {Berezinskii}(1972)}]{Berezinskii_1972}%
  \BibitemOpen
  \bibfield  {author} {\bibinfo {author} {\bibfnamefont {V.~L.}\ \bibnamefont
  {Berezinskii}},\ }\href@noop {} {\bibfield  {journal} {\bibinfo  {journal}
  {Sov. Phys. JETP}\ }\textbf {\bibinfo {volume} {34}},\ \bibinfo {pages} {610}
  (\bibinfo {year} {1972})}\BibitemShut {NoStop}%
\bibitem [{\citenamefont {Kosterlitz}\ and\ \citenamefont
  {Thouless}(1973)}]{Kosterlitz_1973}%
  \BibitemOpen
  \bibfield  {author} {\bibinfo {author} {\bibfnamefont {J.~M.}\ \bibnamefont
  {Kosterlitz}}\ and\ \bibinfo {author} {\bibfnamefont {D.~J.}\ \bibnamefont
  {Thouless}},\ }\href {\doibase 10.1088/0022-3719/6/7/010} {\bibfield
  {journal} {\bibinfo  {journal} {Journal of Physics C: Solid State Physics}\
  }\textbf {\bibinfo {volume} {6}},\ \bibinfo {pages} {1181} (\bibinfo {year}
  {1973})}\BibitemShut {NoStop}%
\bibitem [{\citenamefont {Loktev}\ \emph {et~al.}(2001)\citenamefont {Loktev},
  \citenamefont {Quick},\ and\ \citenamefont {Sharapov}}]{loktevreview}%
  \BibitemOpen
  \bibfield  {author} {\bibinfo {author} {\bibfnamefont {V.~M.}\ \bibnamefont
  {Loktev}}, \bibinfo {author} {\bibfnamefont {R.~M.}\ \bibnamefont {Quick}}, \
  and\ \bibinfo {author} {\bibfnamefont {S.~G.}\ \bibnamefont {Sharapov}},\
  }\href {\doibase 10.1016/S0370-1573(00)00114-9} {\bibfield  {journal}
  {\bibinfo  {journal} {Physics Reports}\ }\textbf {\bibinfo {volume} {349}},\
  \bibinfo {pages} {1} (\bibinfo {year} {2001})}\BibitemShut {NoStop}%
\bibitem [{\citenamefont {Loktev}\ and\ \citenamefont
  {Turkowski}(2009)}]{loktev}%
  \BibitemOpen
  \bibfield  {author} {\bibinfo {author} {\bibfnamefont {V.~M.}\ \bibnamefont
  {Loktev}}\ and\ \bibinfo {author} {\bibfnamefont {V.}~\bibnamefont
  {Turkowski}},\ }\href {\doibase 10.1103/PhysRevB.79.233402} {\bibfield
  {journal} {\bibinfo  {journal} {Phys. Rev. B}\ }\textbf {\bibinfo {volume}
  {79}},\ \bibinfo {pages} {233402} (\bibinfo {year} {2009})}\BibitemShut
  {NoStop}%
\bibitem [{\citenamefont {McMillan}(1968)}]{mcmillan}%
  \BibitemOpen
  \bibfield  {author} {\bibinfo {author} {\bibfnamefont {W.~L.}\ \bibnamefont
  {McMillan}},\ }\href {\doibase 10.1103/PhysRev.167.331} {\bibfield  {journal}
  {\bibinfo  {journal} {Phys. Rev.}\ }\textbf {\bibinfo {volume} {167}},\
  \bibinfo {pages} {331} (\bibinfo {year} {1968})}\BibitemShut {NoStop}%
\bibitem [{\citenamefont {Neilson}\ \emph {et~al.}(2014)\citenamefont
  {Neilson}, \citenamefont {Perali},\ and\ \citenamefont
  {Hamilton}}]{PhysRevB.89.060502}%
  \BibitemOpen
  \bibfield  {author} {\bibinfo {author} {\bibfnamefont {D.}~\bibnamefont
  {Neilson}}, \bibinfo {author} {\bibfnamefont {A.}~\bibnamefont {Perali}}, \
  and\ \bibinfo {author} {\bibfnamefont {A.~R.}\ \bibnamefont {Hamilton}},\
  }\href {\doibase 10.1103/PhysRevB.89.060502} {\bibfield  {journal} {\bibinfo
  {journal} {Phys. Rev. B}\ }\textbf {\bibinfo {volume} {89}},\ \bibinfo
  {pages} {060502} (\bibinfo {year} {2014})}\BibitemShut {NoStop}%
\bibitem [{\citenamefont {Perali}\ \emph {et~al.}(2013)\citenamefont {Perali},
  \citenamefont {Neilson},\ and\ \citenamefont
  {Hamilton}}]{PhysRevLett.110.146803}%
  \BibitemOpen
  \bibfield  {author} {\bibinfo {author} {\bibfnamefont {A.}~\bibnamefont
  {Perali}}, \bibinfo {author} {\bibfnamefont {D.}~\bibnamefont {Neilson}}, \
  and\ \bibinfo {author} {\bibfnamefont {A.~R.}\ \bibnamefont {Hamilton}},\
  }\href {\doibase 10.1103/PhysRevLett.110.146803} {\bibfield  {journal}
  {\bibinfo  {journal} {Phys. Rev. Lett.}\ }\textbf {\bibinfo {volume} {110}},\
  \bibinfo {pages} {146803} (\bibinfo {year} {2013})}\BibitemShut {NoStop}%
\bibitem [{\citenamefont {Debnath}\ \emph {et~al.}(2017)\citenamefont
  {Debnath}, \citenamefont {Barlas}, \citenamefont {Wickramaratne},
  \citenamefont {Neupane},\ and\ \citenamefont {Lake}}]{PhysRevB.96.174504}%
  \BibitemOpen
  \bibfield  {author} {\bibinfo {author} {\bibfnamefont {B.}~\bibnamefont
  {Debnath}}, \bibinfo {author} {\bibfnamefont {Y.}~\bibnamefont {Barlas}},
  \bibinfo {author} {\bibfnamefont {D.}~\bibnamefont {Wickramaratne}}, \bibinfo
  {author} {\bibfnamefont {M.~R.}\ \bibnamefont {Neupane}}, \ and\ \bibinfo
  {author} {\bibfnamefont {R.~K.}\ \bibnamefont {Lake}},\ }\href {\doibase
  10.1103/PhysRevB.96.174504} {\bibfield  {journal} {\bibinfo  {journal} {Phys.
  Rev. B}\ }\textbf {\bibinfo {volume} {96}},\ \bibinfo {pages} {174504}
  (\bibinfo {year} {2017})}\BibitemShut {NoStop}%
\bibitem [{\citenamefont {Wang}\ \emph
  {et~al.}(2019{\natexlab{b}})\citenamefont {Wang}, \citenamefont {Rhodes},
  \citenamefont {Watanabe}, \citenamefont {Taniguchi}, \citenamefont {Hone},
  \citenamefont {Shan},\ and\ \citenamefont {Mak}}]{WangNature2019}%
  \BibitemOpen
  \bibfield  {author} {\bibinfo {author} {\bibfnamefont {Z.}~\bibnamefont
  {Wang}}, \bibinfo {author} {\bibfnamefont {D.~A.}\ \bibnamefont {Rhodes}},
  \bibinfo {author} {\bibfnamefont {K.}~\bibnamefont {Watanabe}}, \bibinfo
  {author} {\bibfnamefont {T.}~\bibnamefont {Taniguchi}}, \bibinfo {author}
  {\bibfnamefont {J.~C.}\ \bibnamefont {Hone}}, \bibinfo {author}
  {\bibfnamefont {J.}~\bibnamefont {Shan}}, \ and\ \bibinfo {author}
  {\bibfnamefont {K.~F.}\ \bibnamefont {Mak}},\ }\href {\doibase
  10.1038/s41586-019-1591-7} {\bibfield  {journal} {\bibinfo  {journal} {Nature
  (London)}\ }\textbf {\bibinfo {volume} {574}},\ \bibinfo {pages} {76}
  (\bibinfo {year} {2019}{\natexlab{b}})}\BibitemShut {NoStop}%
\bibitem [{\citenamefont {Bagnato}\ and\ \citenamefont
  {Kleppner}(1991)}]{Bagnato1991}%
  \BibitemOpen
  \bibfield  {author} {\bibinfo {author} {\bibfnamefont {V.}~\bibnamefont
  {Bagnato}}\ and\ \bibinfo {author} {\bibfnamefont {D.}~\bibnamefont
  {Kleppner}},\ }\href {\doibase 10.1103/PhysRevA.44.7439} {\bibfield
  {journal} {\bibinfo  {journal} {Phys. Rev. A}\ }\textbf {\bibinfo {volume}
  {44}},\ \bibinfo {pages} {7439} (\bibinfo {year} {1991})}\BibitemShut
  {NoStop}%
\bibitem [{\citenamefont {Hohenberg}(1967)}]{Hohenberg1967}%
  \BibitemOpen
  \bibfield  {author} {\bibinfo {author} {\bibfnamefont {P.~C.}\ \bibnamefont
  {Hohenberg}},\ }\href {\doibase 10.1103/PhysRev.158.383} {\bibfield
  {journal} {\bibinfo  {journal} {Phys. Rev.}\ }\textbf {\bibinfo {volume}
  {158}},\ \bibinfo {pages} {383} (\bibinfo {year} {1967})}\BibitemShut
  {NoStop}%
\end{thebibliography}%
\bibliographystyle{apsrev4-1}

\end{document}